\documentclass[aps,preprint,nofootinbib,amsmath,11pt]{revtex4}

\usepackage{verbatim}
\usepackage{yfonts}
\usepackage{bm}
\usepackage{epsfig}
\usepackage{psfrag}

\def\Nc{N_{\rm c}}
\def\k{{\bm k}}
\def\CA{{\cal A}}
\def\CB{{\cal B}}
\newcommand{\qn}{{\textswab{q}}}
\newcommand{\wn}{{\textswab{w}}}

\begin{document}

\title{Thermal spectral functions of strongly coupled
${\cal N}{=}4$
supersymmetric Yang-Mills theory}

\author{Pavel Kovtun}
\email{kovtun@kitp.ucsb.edu}
\affiliation{KITP, University of California, Santa Barbara, CA 93106, USA}%
\author{Andrei Starinets}
\email{starina@perimeterinstitute.ca}
\affiliation{Perimeter Institute for Theoretical Physics,
  Waterloo, ON N2L 2Y5, Canada}

\date{February 2006\\[30pt]}

\begin{abstract}
\noindent
We use the gauge-gravity duality conjecture to compute
spectral functions of the stress-energy tensor
in finite temperature ${\cal N}{=}4$ supersymmetric Yang-Mills theory
in the limit of large $\Nc$ and large 't~Hooft coupling.
The spectral functions exhibit peaks characteristic of
hydrodynamic modes at small frequency, and oscillations
at intermediate frequency.
The non-perturbative spectral functions differ qualitatively from those
obtained in perturbation theory.
The results may prove useful for lattice studies of 
transport processes in thermal gauge theories.
\end{abstract}

\preprint {NSF-KITP-06-06}
\maketitle

%
%

\section{Introduction}
Theoretical analysis of the properties of
strongly interacting hot and dense matter is a hard problem.
Even when the density of baryons in thermal equilibrium is
negligible, perturbative QCD calculations are only reliable
at temperatures which are much higher than the temperature of the
deconfinement transition~\cite{talks}.
While lattice simulations can provide an account of equilibrium
thermodynamic properties of the theory, questions
involving real-time dynamics such as momentum transport,
thermalization, and various production rates are much harder to answer.
For near-equilibrium states, physical information is contained
in equilibrium response functions:
for example, the dilepton rate is proportional to the spectral function
of vector currents, and the viscosities are determined by the
spectral function of the relevant components of the stress-energy tensor.

Thus it is valuable to study
models where analytic results for the real-time response functions
can be derived.
A number of recent studies focused on a particular model, the
${\cal N}{=}4$ supersymmetric $SU(\Nc)$ Yang-Mills (SYM)
theory at finite temperature.
The interest in this theory 
is due to Maldacena's gauge-string duality conjecture \cite{agmoo}
which provides an effective description 
of the theory's non-perturbative regime in terms 
of semiclassical gravity in a five-dimensional 
asymptotically Anti de Sitter space.

The SYM theory is conformal,
and has only one tunable parameter, the 't~Hooft coupling $\lambda$.
Thermal equilibrium is characterized by the black-body equation of state
at any non-zero temperature, and can not
(at large 't~Hooft coupling)
be viewed as a gas of weakly interacting quasiparticles.
The standard hydrodynamic singularities in two-point functions
of conserved currents in SYM theory were found
from the dual gravity description in the limit
of large $\lambda$ and large $\Nc$ \cite{PSS-shear,PSS-sound}.
The ratio of the shear viscosity to volume entropy density in this theory
has been found to be
$\eta/s={1}/{4\pi}$ 
\cite{PSS-viscosity, sochinenie},
which is a much smaller number than the corresponding result in a
weakly coupled gauge theory.

In this paper, we focus on the full spectral functions
(rather than their small-frequency limit)
of the stress-energy tensor in the SYM theory
at large 't~Hooft coupling and large $\Nc$.
The  knowledge of the full 
spectral function is important for two reasons.
On the one hand, our results provide the first example
of a non-perturbative spectral function calculation in a strongly coupled
four-dimensional gauge theory at finite temperature,
obtained without performing lattice simulations.
On the other hand, the non-perturbative spectral functions of
SYM theory
may be useful for lattice computations of transport coefficients
in realistic gauge theories at temperatures not too far from the
deconfinement transition.
Indeed, a Euclidean correlation function (computed on the lattice)
is proportional to the integral of the real-time spectral function
over all frequencies.
In the simplest method of reconstructing the
spectral function from the Euclidean data, one first assumes
an ansatz for the spectral function, and then fits the parameters of the 
ansatz to the lattice data \cite{Karsch:1986cq}.
This method has been applied to the computation of
the shear viscosity in QCD \cite{Nakamura-Sakai}.
Theoretical understanding of what
the correct non-perturbative ansatz might be is of primary
importance for this approach.
Our results for the non-perturbative spectral functions of
SYM theory may therefore prove useful for lattice studies
of transport processes in thermal gauge theories.


\section{Correlation functions from gauge-gravity duality}

The spectral function $\chi_{\mu\nu,\alpha\beta}(k)$ is defined as
\begin{equation}
  \chi_{\mu\nu,\alpha\beta}(k) = 
  \int d^4 x\; e^{-ikx} \langle [T_{\mu\nu}(x), T_{\alpha\beta}(0)]\rangle\ .
\end{equation}
It is proportional to the imaginary part of the retarded Green's function,
$
\chi_{\mu\nu,\alpha\beta}(k)=-2\, {\rm Im}\, G_{\mu\nu,\alpha\beta}(k)\,,
$
where
\begin{equation}
  G_{\mu\nu, \alpha\beta} (k)
  = -i\!\int\!d^4x\,e^{-i k x}\,
  \theta(x^0) \langle
 [T_{\mu\nu}(x),\, T_{\alpha\beta }(0)] \rangle \,,
\label{retarded}
\end{equation}
and is an odd, real function of $k$.
The shear and bulk viscosities are
proportional to the zero-frequency slope
of the specific components of the spectral function,
for example,
\begin{eqnarray}
   \eta &=& 
   \lim_{k^0\to0}\frac{1}{2k^0}\,\chi_{xy,xy}(k^0,\k{=}0)\,.
   \label{eq:Kubo-formula}
\end{eqnarray}
The retarded  correlation function
of the stress-energy tensor of
SYM theory admits a simple decomposition
(we follow the notational conventions of Ref.~\cite{quasipaper}).
At zero temperature, Lorentz symmetry combined with
conservation and tracelessness of $T_{\mu\nu}$ implies that
$G_{\mu\nu,\alpha\beta}(k)$
has the form
$$G_{\mu\nu,\alpha\beta}(k)=H_{\mu\nu,\alpha\beta}\, G_S(k^2)\ ,$$
where 
$
    H_{\mu\nu,\alpha\beta} =
       \frac12\left(P_{\mu\alpha}P_{\nu\beta}+
       P_{\mu\beta}P_{\nu\alpha}\right)-
       \frac{1}{3}P_{\mu\nu}P_{\alpha\beta}
$
is a projector onto conserved traceless symmetric tensors,
$P_{\mu\nu}=\eta_{\mu\nu}-k_\mu k_\nu/k^2$, and $k^2=-(k^0)^2+\k^2$.
At non-zero temperature, $G_{\mu\nu,\alpha\beta}$
in a conformal theory
can be described by three symmetry channels
\begin{equation}
   G_{\mu\nu,\alpha\beta}(k) =
   S_{\mu\nu,\alpha\beta}\, G_1 +
   Q_{\mu\nu,\alpha\beta}\, G_2 +
   L_{\mu\nu,\alpha\beta}\, G_3 \,,
\label{eq:three-channels}
\end{equation}
where
$S_{\mu\nu,\alpha\beta}$, $Q_{\mu\nu,\alpha\beta}$,
$L_{\mu\nu,\alpha\beta}$ are the appropriate orthogonal projectors
which provide three independent Lorentz index structures~\cite{quasipaper}.
Choosing the spatial momentum along $x^3$, $k_\mu=(-\omega,0,0,q)$,
the components of the correlation function are
\begin{subequations}
\begin{eqnarray}
  &&G_{tx^1,tx^1}(k)
      = \frac12 \frac{q^2}{\omega^2{-}q^2} G_1(\omega,q) ,
 \label{eq:Gtxtx}\\
  &&G_{tt,tt}(k)= \frac23 \frac{q^4}{(\omega^2{-}q^2)^2} G_2(\omega,q)\ ,
 \label{eq:Gtttt}\\
  &&G_{x^1 x^2,x^1 x^2}(k) = \frac12 G_3(\omega,q)\,,
  \label{eq:Gscal} 
\end{eqnarray}
\end{subequations}
with all other related to the above by the rotation invariance.
As a function of complex $\omega$,
in the low-frequency limit, $G_1(\omega,q)$
has a shear-mode singularity,
$G_2(\omega,q)$ has a sound-mode singularity,
and $G_3(\omega,q)$ has no hydrodynamic singularities.
In the limit of vanishing 3-momentum,
$G_1(\omega)=G_2(\omega)=G_3(\omega)$.
At zero temperature, $G_1=G_2=G_3=G_S$.

In the regime of large 't~Hooft coupling,
the three scalar functions
$G_a(\omega,q)$ can be computed using the gauge-gravity duality recipe
\cite{PSS-recipe,quasipaper}.
The duality essentially reduces the computation
of a two-point correlation function to a boundary-value problem
for a linear ordinary differential equation.
For the zero-temperature theory, the retarded two-point function
can be found for example in \cite{PSS-recipe}%
\footnote{
	At zero temperature, the dual gravity result for the correlator
	in strongly coupled ${\cal N}{=}4$ SYM theory 
	coincides with the one obtained in free field theory, due to 
	a non-renormalization theorem \cite{Gubser-Klebanov}.},
\begin{equation}
   G_S(k)=\frac{\Nc^2 k^4}{32\pi^2} \left( 
         \ln|k^2| -i\pi\theta(-k^2){\rm sign}\, \omega   \right)\,.
\label{eq:zeroT-G}
\end{equation}
In order to compute the retarded correlators at non-zero temperature,
one has to analyze the ``wave equations'' (one for each
symmetry channel) which describe propagation of the corresponding
metric perturbations in the
AdS-Schwarzschild background spacetime of the dual description.
The differential equations are of the form
\begin{equation}
  \frac{d^2}{du^2} Z_a(u) +
  p_a(u) \frac{d}{du}Z_a(u) +
  q_a(u) Z_a(u) = 0 \,,
\label{eq:master-equation}
\end{equation}
where the coefficients $p_a(u)$, $q_a(u)$
[to be specified shortly]
depend on the dimensionless frequency $\wn\equiv\omega/2\pi T$
and momentum $\qn\equiv q/2\pi T$,
and $a=1,2,3$ labels the three symmetry channels.
The coordinate $u$ ranges from $0$ to $1$,
where $u=0$ corresponds to the boundary of the 
asymptotically AdS spacetime, and
$u=1$ corresponds to the event horizon of the background.

For all three equations (\ref{eq:master-equation}), the characteristic
exponents at  $u=0$ are equal to $0$ and $2$,  and
the exponents at $u=1$ are $\pm i\wn/2$.
Information about the retarded correlation function
is encoded in the solutions to Eq.~(\ref{eq:master-equation})
which satisfy the incoming wave condition
at the horizon,
corresponding to the exponent $-i\wn/2$ at $u=1$.
The correct solution is thus of the form
$Z_a(u) = (1-u)^{-i\wn/2} F_a(u)$, where $F_a(u)$ is a regular
function at the horizon. 
The solution satisfying the incoming-wave condition
at the horizon can be written  as a
linear combination of two independent local solutions at $u=0$,
\begin{equation}
  Z_a(u) = \CA_{a} Z_a^{I}(u) + \CB_{a} Z_a^{II}(u) \,,
\label{eq:Z-two-solutions}
\end{equation}
where $Z_a^{I}(u)$ and  $Z_a^{II}(u)$ are given by their standard 
Frobenius expansions \cite{yellow-book} as 
{\setlength\arraycolsep{2pt}
\begin{eqnarray}
&&\!\!\!\!\!\!\!\!
  Z_a^{I} = 1+ b_{aI}^{(1)}\, u+ 
             h_{a}\,  Z_a^{II}(u)\, \ln{u} + b_{aI}^{(2)}\, u^2+ \cdots\,,
  \label{frob1} \\
&&\!\!\!\!\!\!\!\!
  Z_a^{II} = u^2 \! \left( 
  1+ b_{aI\!I}^{(1)}\, u+ b_{aI\!I}^{(2)}\, u^2+ \cdots \right)\,.
\label{frob2}
\end{eqnarray}}
All the coefficients $b_{aI,I\!I}^{(j)}$ (except $b_{aI}^{(2)}$)
and $h_a$ are determined by the recursion relations obtained by 
substituting the above expansion in the
differential equation (\ref{eq:master-equation}). The coefficient 
 $b_{aI}^{(2)}$  is left as a free parameter, reflecting the fact that  
one can always redefine the local solutions
by adding a constant multiple of  $Z_a^{II}(u)$ to  $Z_a^{I}(u)$.
Without loss of generality, we set  $b_{aI}^{(2)}=0$ thus 
fixing the definition of $Z_a^{I, II}(u)$.
The retarded functions $G_a$ are then given by%
\footnote{
	The real parts of the correlator in the shear and sound channels
	have additional contributions arising from writing the 
	gravitational action in terms of the 
	gauge-invariant variables \cite{quasipaper}.
	These terms are irrelevant for the present discussion.
}
\begin{equation}
   G_a(\omega,q) = -\pi^2 \Nc^2 T^4 
    \frac{\CB_{a}(\omega,q)}{\CA_{a}(\omega,q)}\,.
\end{equation}
As is evident from Eqs.~(\ref{frob1}),  (\ref{frob2}),
the coefficient $\CA$ is given by the boundary value
of the solution, $\CA_{a}=\lim_{u\to0}Z_a(u)$, while the coefficient 
$\CB$ can be expressed in terms of the
boundary value of the second derivative of the solution,
\begin{equation}
  \CB_a = \frac12 \lim_{u\to0}\left( Z_a''(u)- 2\CA_a h_a \ln(u)\right) -
         \frac32  \CA_a  h_a \,.
\end{equation}
From the recursion relations one determines $h_a=-\frac12(\qn^2-\wn^2)^2$,
which is an analytic function of $\wn$, $\qn$, and therefore
represents a contact term which we drop.
The retarded correlator is therefore equal to
\begin{equation}
  G_a = -\pi^2 \Nc^2 T^4 
  \lim_{u\to0} \left(\frac{Z_a''(u)}{2 Z_a(u)}- h_a\ln(u)\right) \,.
\end{equation}

\section{Spectral functions of the stress-energy tensor}

\begin{figure}
  \psfrag{w}{\!\!\wn}
  \includegraphics[width=3.0in]{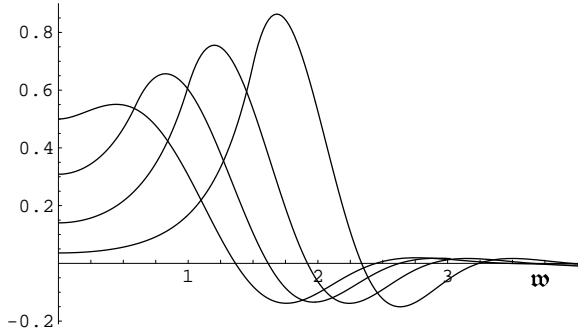}
\caption{
	Finite-temperature part of the spectral function 
	for transverse stress
	$(\chi_{xy,xy}-\chi_{xy,xy}^{T=0})/\wn$,
	plotted in units of $\pi^2\Nc^2 T^4$ as a function of
	dimensionless frequency $\wn\equiv\omega/2\pi T$.
	Different curves correspond to values of the
	dimensionless spatial momentum $\qn\equiv q/2\pi T$
	equal to $0$, $0.6$, $1.0$, and $1.5$.
}
\label{fig:scalar}
\end{figure}

In order to determine the spectral function
for transverse stress,
we need to solve the master equation
(\ref{eq:master-equation}) 
whose coefficients are given by \cite{PSS-recipe}
\begin{equation}
  p_3(u) = -\frac{1+u^2}{uf}\ , \ \ \ \ 
  q_3(u) = \frac{\wn^2 - \qn^2 f}{u f^2}\ ,
\end{equation}
where $f=1-u^2$.  
Solving Eq.~(\ref{eq:master-equation}) numerically, 
we find $\chi_{xy,xy}(\omega,q)$ 
as explained in the previous section.
The spectral function is linear in $\wn$ for small frequencies, 
$\chi_{xy,xy}=(\wn/2) \pi^2\Nc^2 T^4 (1+O(\wn^2,\qn^2))$,
then increases monotonically, and
at large frequencies
it asymptotes to the zero-temperature result
$\chi_{xy,xy}^{T=0}=\frac\pi2 (\qn^2-\wn^2)^2\, \pi^2 \Nc^2 T^4 \theta(-k^2)$.
When this zero-temperature contribution is subtracted,
the resulting function exhibits oscillations which damp rapidly
as frequency grows.
The oscillations appear around $\omega=q$,
and their amplitude grows with $q$.
The unsubtracted $\chi_{xy,xy}$
is positive, as it should be.
Figure~\ref{fig:scalar} shows graphs of
$\chi_{xy,xy}$ for several values of three-momentum%
\footnote{
	The numerical procedure outlined above allows one to compute
	both real and imaginary parts of $G_{\mu\nu,\alpha\beta}$.
	The real part which can
	in principle be reconstructed from the spectral function,
	also exhibits oscillations which damp
	as frequency grows~\cite{Hartnoll-Kumar}.
}.
The shear viscosity follows from  the Kubo formula 
(\ref{eq:Kubo-formula}), $\eta=\pi\Nc^2 T^3/8$, and is 
in agreement with the earlier results~\cite{PSS-viscosity,sochinenie}.

\begin{figure}
\begin{minipage}[t]{0.48\textwidth}
  \psfrag{1.4}{\ \ \wn}
  \includegraphics[width=3.0in]{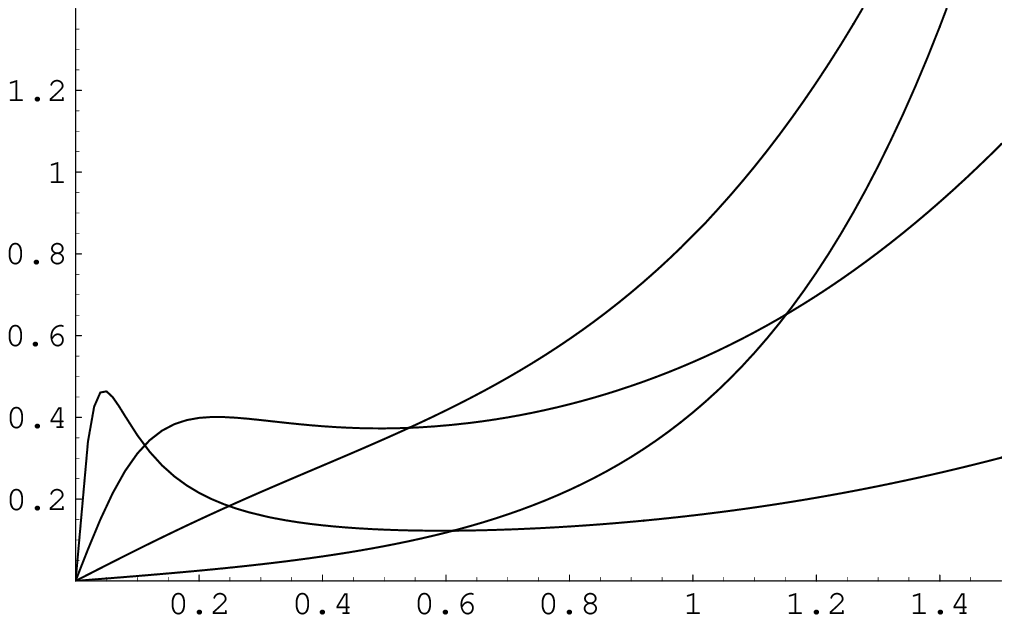}
\end{minipage}
\begin{minipage}[t]{0.48\textwidth}
  \psfrag{w}{\ \ \ \wn}
  \includegraphics[width=3.0in]{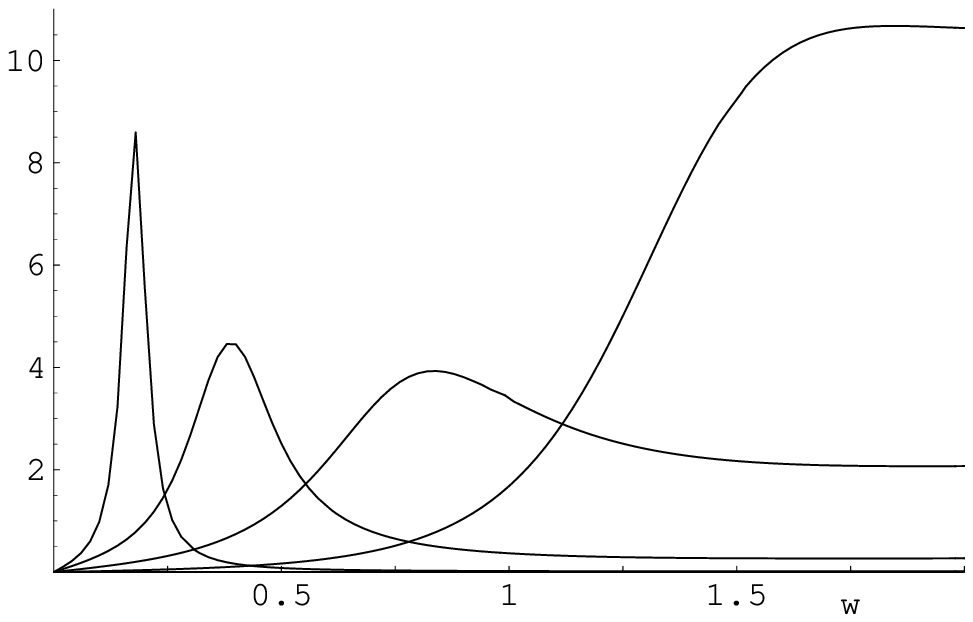}
\end{minipage}
\caption{
	Left: spectral function for longitudinal momentum density,
	$\chi_{tx,tx}$,
	plotted in units of $\pi^2\Nc^2 T^4$, as a function of
	dimensionless frequency $\wn\equiv\omega/2\pi T$.
	Different curves correspond to values of the
	dimensionless spatial momentum $\qn\equiv q/2\pi T$
	equal to $0.3$, $0.6$, $1.0$, and $1.5$.
	At large $\wn$, the curves asymptote to
	the zero-temperature result $\frac\pi2 \qn^2 (\wn^2-\qn^2)$.
	Right: spectral function for energy density,
	$\chi_{tt,tt}$,
	plotted in units of $\pi^2\Nc^2 T^4$, as a function of
	dimensionless frequency $\wn\equiv\omega/2\pi T$.
	Different curves correspond to values of the
	dimensionless spatial momentum $\qn\equiv q/2\pi T$
	equal to $0.3$, $0.6$, $1.0$, and $1.5$.
	At large $\wn$, the curves asymptote to
	the zero-temperature result $\frac\pi2\, 4\qn^4/3$.
}
\label{fig:shear-sound}
\end{figure}

In the shear channel,
the coefficients of 
the master equation (\ref{eq:master-equation})
are given by~\cite{quasipaper}
\begin{equation}
  p_1(u){=}\frac{(\wn^2-\qn^2 f)f + 2 u^2\wn^2}{uf(\qn^2 f-\wn^2)}\,, \ \  
  q_1(u){=}\frac{\wn^2 - \qn^2 f}{uf^2}\, .
\end{equation}
The spectral function $\chi_{tx,tx}(\omega,q)$ is shown in 
Fig.~\ref{fig:shear-sound}. 
At small momentum, the spectral function exhibits a narrow peak at
$\wn=\qn^2/2$, characteristic of the hydrodynamic shear mode.

In the sound channel,
the coefficients 
of the master equation (\ref{eq:master-equation})
are given by \cite{quasipaper}
{\setlength\arraycolsep{1pt}
\begin{eqnarray}
  p_2(u) &=& -\frac{3\wn^2 (1+u^2) + \qn^2 ( 2u^2 - 3 u^4 -3)}
           {u f (3 \wn^2 +\qn^2 (u^2-3))}\ , \\
  q_2(u) &=&  \frac{3 \wn^4 +\qn^4 (3{-}4u^2{+}u^4) +
            \qn^2 (4u^2\wn^2{-}6\wn^2{-}4u^3 f)}
            {u f^2 ( 3 \wn^2 + \qn^2 (u^2 -3))}\, . 
\end{eqnarray}}
The spectral function  $\chi_{tt,tt}(\omega,q)$ is shown 
in Fig.~\ref{fig:shear-sound}.
At small momenta, the spectral function exhibits a narrow peak at
$\wn=\qn/\sqrt{3}$, characteristic of the hydrodynamic sound mode.
For both $\chi_{tx,tx}$ and $\chi_{tt,tt}$,
the finite-temperature contributions have oscillatory behavior,
similar to the one seen in $\chi_{xy,xy}$.

\section{Discussion}
It is easy to understand the limiting behavior
of the spectral functions found above.
The $\omega$, $q$ dependence at small frequency is predicted by
the linearized hydrodynamics, while the large $\omega$ dependence is
fixed by the scale invariance of the SYM theory.
An intriguing  feature is the presence of oscillations in the
finite-temperature contribution, which appear around $\omega=q$,
and then decay rapidly.
Mathematically, such damped oscillations are due to the characteristic 
asymptotic behavior, $\sim \exp{(-\alpha \wn)}$, 
of the solutions to  Eq.~(\ref{eq:master-equation}), 
where $\alpha$ is a complex number 
\cite{Policastro:2001yb,PSS-recipe}. 
They reflect the presence of an infinite sequence of poles
in the lower half-plane of the retarded correlators~\cite{quasipaper}.

One can compare the spectral functions of the 
strongly coupled SYM theory
with the perturbative results in a weakly coupled scalar
or pure gauge theory discussed in Ref.~\cite{Aarts-Resco}.
At weak coupling, the spectral function for transverse stress
at $q{=}0$ grows linearly at small frequency, and is proportional to 
$\omega^4$
at asymptotically high frequency.
In between, however, there is a range where the spectral function
decreases as $1/\omega$.
This behavior at intermediate frequencies at weak coupling
differs from what one observes in the strongly coupled SYM theory,
where the spectral function grows monotonically for all $\omega$.
Based on the numerical results for $\chi_{\mu\nu,\alpha\beta}(\omega,q)$,
one may propose the following
ansatz for the zero-momemtum spectral function
for transverse stress
\begin{equation}
   \chi_{xy,xy}(\omega)=\chi_{xy,xy}^{T=0}(\omega) +
   {\rm Im} \sum_i c_i\, e^{-\alpha_i\, \omega} \,,
\label{eq:SYM-ansatz}
\end{equation}
where $c_i$, are real coefficients,
and ${\rm Re}(\alpha_i)>0$ (assuming positive frequency).

In this paper, we computed the spectral functions 
in the simplest thermal gauge theory with a known gravity dual.
It should also be feasible to find spectral functions for a
class of non-conformal gauge theories, such as the one
analyzed in Ref.~\cite{Benincasa:2005iv}.

{\it Note added:} while this paper was being completed,
we became aware of the work on the same subject
by D.~Teaney~\cite{Teaney}.

\begin{acknowledgments}
We are grateful to L.G.~Yaffe for helpful discussions.
The work of P.K. was supported in part by the
National Science Foundation under Grant No. PHY99-07949.
Research at Perimeter institute is supported in part
by funds from NSERC of Canada.
P.K. acknowledges the hospitality of Perimeter Institute
where part of this work was completed.
\end{acknowledgments}

\end{document}